\documentclass[12pt,showpacs]{iopart}
\usepackage{epsfig}
\usepackage{amsbsy}
\usepackage{amsmath}
\usepackage{amsfonts}

\begin{document}

\newcommand{\nn}{\nonumber}
\newcommand{\bra}[1]{\langle{#1}|}
\newcommand{\ket}[1]{|{#1}\rangle}
\newcommand{\pra}{{\it Phys. Rev. A }}
\newcommand{\nl}{\\[0.3 cm]}
\newcommand{\hl}{\\[-0.2 cm]}

\title{Bounds on general entropy measures}
\author{Dominic W Berry and Barry C Sanders\footnote{Present address:
Quantum Information Science Group,
Department of Physics and Astronomy, University of Calgary,
Calgary, Alberta T2N 1N4, Canada.}}
\address{Australian Centre for Quantum Computer Technology,
Department of Physics, \\
Macquarie University, Sydney, New South Wales 2109, Australia}
\date{\today}

\begin{abstract}
We show how to determine the maximum and minimum possible values of one measure
of entropy for a given value of another measure of entropy. These maximum
and minimum values are obtained for two standard forms of probability
distribution (or quantum state) independent of the entropy measures, provided the
entropy measures satisfy a concavity/convexity relation. These results
may be applied to entropies for classical probability distributions, entropies
of mixed quantum states and measures of entanglement for pure states.
\end{abstract}
\maketitle

\section{Introduction}
Entropy plays a significant role in both classical and quantum information
theories and is characterised by various measures, including the Shannon
entropy in classical information theory and the von Neumann entropy for a
density operator of a quantum system. These entropy measures are related,
because the von Neumann entropy of a density operator equals the Shannon
entropy of the eigenvalues of this operator. The von Neumann and Shannon
entropies are the standard entropy measures, but other entropy
measures are also employed.

For quantum systems, the linear entropy is widely employed. The linear
entropy is easier to calculate than the von Neumann entropy in general,
hence its appeal. Other entropy measures are also employed,
including the Tsallis entropy \cite{Tsallis} and the $\alpha$, or R\'{e}nyi,
entropy \cite{Renyi}. A common way of describing these entropy measures is
via the trace over a concave function of the density operator. This general
entropy measure has been widely studied both in the context of classical
probability distributions and mixed quantum states \cite{general}.

This work is motivated by problems where one entropy measure is required,
but it is only possible to obtain analytic results for another entropy
measure. However, the exact result for the desired entropy measure may
not be required. Driven by this motivation, we establish a method by which
one entropy measure can be estimated from another entropy measure that may
be easier to calculate. Specific examples of this problem have been
considered in \cite{XKK,Harre,Wei}; here we derive the general
result for general entropy measures in both the classical and quantum
contexts.

General entropy measures for classical information and for quantum
information are discussed in section \ref{sec:general}. In
section \ref{sec:bounds}, we show that the states and probability distributions
given in \cite{Harre,Wei} extremise (minimise or maximise) general
entropy measures for a given value of another generalised entropy provided
that a concavity/convexity condition is satisfied. In section \ref{sec:apply},
we apply our methods to particular examples and provide an example of an
entropy measure that violates the concavity/convexity condition. We conclude
in section \ref{sec:conclude}.

\section{General measures of entropy}
\label{sec:general}
Three common measures of quantum entropy are the von Neumann, linear and R\'enyi
entropies. The von Neumann entropy for density operator $\rho$ is given by
$S_{\rm vN}(\rho)\equiv-{\rm Tr}(\rho\log\rho)$,
where the notation $\log$ is used for logarithms base $2$.
The linear entropy is defined by
$S_{\rm lin}(\rho)\equiv 1-{\rm Tr}(\rho^2)$,
and the $\alpha$ or R\'enyi entropy \cite{Renyi} by
\begin{equation}
S_{\alpha}(\rho)\equiv\frac 1{1-\alpha}\log{\rm Tr}(\rho^\alpha).
\end{equation}
These three entropy measures are all calculated from an expression
of the form ${\rm Tr}F(\rho)$.

In order to derive general results, we therefore consider general entropy
measures of the form \cite{general}
\begin{equation}
S_f(\rho) \equiv {\rm Tr}F(\rho).
\end{equation}
The function $f$ is a mapping $[0,1]\mapsto\mathbb{R}$, and $F$ is the
corresponding operator function defined by
\begin{equation}
F(\rho) \equiv \sum_{i=0}^{d-1} f(\lambda_i) \ket{\phi_i}\bra{\phi_i}
\end{equation}
where $\lambda_i$ and $\ket{\phi_i}$ are the eigenvalues and eigenstates,
respectively, of $\rho$, and $d$ is the dimension of the
Hilbert space. The entropy measure $S_f$ therefore only depends on the
eigenvalues of the density matrix, and may be calculated as
$S_f(\rho) = \sum_i f(\lambda_i)$.
We require that the function $f$ satisfies the following three conditions: \\ \\
{\it Condition 1}: $f(0)=0$. \\
{\it Condition 2}: The function $f$ is strictly concave or strictly convex. \\
{\it Condition 3}: The first derivative $f'$ exists and is continuous in the
interval $(0,1)$. \\

All the examples of entropy measures above satisfy these three conditions.
The first condition allows us to embed the Hilbert space in another of larger
dimension without changing the value obtained for $S_f$. The second condition
implies that the extremal values of the entropy are obtained for pure
and maximally mixed states. This result follows from the following lemma: \nl
{\bf Lemma 1.} {\it Let $f:[0,1]\to\mathbb{R}$ be a function such that $f(0)=0$,
and let $\{\lambda_i\}$ be a set of $d$ non-negative real numbers such that
$\sum_i\lambda_i=\Lambda \le 1$. If $f$ is concave (convex), then the minimum
(maximum) value of $\sum_i f(\lambda_i)$ is obtained for one of the
$\lambda_i$ equal to $\Lambda$ and the rest zero, and the maximum (minimum)
value of $\sum_i f(\lambda_i)$ is obtained for all $\lambda_i$ equal.} \nl
{\bf Proof.} This result is well known; however, we show the result here for
completeness. Concavity (convexity) implies
\begin{equation}
\sum_i f(\lambda_i) \begin{array}{*{20}c} \ge \\ (\le) \\ \end{array}
\sum_i \left[ (1-\lambda_i/\Lambda)f(0)+(\lambda_i/\Lambda)f(\Lambda) \right] = f(\Lambda).
\end{equation}
Thus the minimum (maximum) value of $\sum_i f(\lambda_i)$ is obtained for one
of the $\lambda_i$ equal to $\Lambda$ and the rest zero. Similarly concavity
(convexity) implies
\begin{equation}
\sum_i f(\lambda_i) \begin{array}{*{20}c} \le \\ (\ge) \\ \end{array} d\times f(\Lambda/d)
\end{equation}
so the maximum (minimum) value of $\sum_i f(\lambda_i)$ is obtained for all $\lambda_i$
equal to $\Lambda/d$. \hfill $\Box$ \hl

Note that, because we allow the possibility that $f$ is strictly convex, our
measure may be considered to be a measure of entropy or of purity (for simplicity
we always call it entropy). This generality is useful because it allows us to
easily apply our results to cases such as the linear entropy and the $\alpha$
entropy. The third condition is not absolutely necessary for $S_f(\rho)$ to
be a reasonable entropy measure.
However, we include it because it is necessary in order to derive the
bounds in the next section. Note that we do not require that the derivative
exists at the endpoints 0 and 1. Allowing the possibility of derivatives that
diverge at the endpoints means that our results may be applied to the von Neumann
entropy. Note also that conditions 2 and 3 imply that the derivative $f'$ must be
one-to-one. The derivative $f'$ will be monotonically increasing (decreasing)
if $f$ is convex (concave).

Because the entropy measure $S_f$ depends only on the eigenvalues of
the density operator, it is equivalent to the entropy measure for classical
probabilities:
\begin{equation}
H_f(\{p_i\})\equiv\sum_{i=0}^{d-1} f(p_i)
\end{equation}
where $f$ satisfies conditions 1--3.
It is clear that $S_f(\rho)=H_f(\{\lambda_i\})$, where $\{\lambda_i\}$ is the
set of eigenvalues for $\rho$. This is a generalisation of the relation
between the von Neumann entropy and the Shannon entropy. We may also define an
analogous general measure of entanglement for pure states:
\begin{equation}
E_f(\ket{\psi}) \equiv S_f({\rm Tr}_A\ket{\psi}\bra{\psi}) = \sum_{i=0}^{d-1}
f(\lambda_i)
\end{equation}
where $f$ is a function satisfying conditions 1--3, $\ket{\psi}$ is a pure
state shared between two subsystems $A$ and $B$, and $\lambda_i$ are the
Schmidt coefficients of $\ket{\psi}$.

The bounds that we derive in the next section may be applied to all three cases:
$H_f$ for classical entropy, $S_f$ for the entropy of mixed quantum states and
$E_f$ for entanglement of pure quantum states. These three cases are
mathematically identical, although the physical interpretations are different.

\section{Bounds on entropy measures}
\label{sec:bounds}
In this section we show how to determine the upper and lower bounds on one
generalised entropy for a given value of another generalised entropy. We will
present the derivation in terms of the entropy for probability distributions.
The results for entropies of mixed states and entanglement measures
immediately follow from this result.

From \cite{Harre} the maximum and minimum Shannon entropies for a given
value of the index of coincidence (where $g(\lambda)=\lambda^2$) are
obtained for probability distributions
\begin{equation}
\label{maxer}
\{\lambda_i\} = \{\lambda_0,\lambda_1,\cdots,\lambda_1\}
\end{equation}
where $\lambda_1=(1-\lambda_0)/(d-1)\le\lambda_0$ and
\begin{equation}
\label{miner}
\{\lambda_i\} = \{\lambda_0,\cdots,\lambda_0,\lambda_1,0,\cdots,0\}
\end{equation}
where $\lambda_1=1-k\lambda_0<\lambda_0$ and there are
$k=\lfloor 1/\lambda_0\rfloor$ probabilities equal to $\lambda_0$. Note that
both probability distributions \eqref{maxer} and \eqref{miner} are
parametrised by the single real number $\lambda_0$. 
The result given in \cite{Wei} for the von Neumann and linear entropies is
equivalent, except that the coefficients $\lambda_i$ are eigenvalues of a density
matrix.

We provide a proof that these two probability distributions give the
bounds when comparing general entropy measures. The specific result is given
below. \\ \\
{\bf Theorem 1.}
{\it Let $H_g=\sum_i g(\lambda_i)$ and $H_f=\sum_i f(\lambda_i)$ be two entropy
measures where the functions $f$ and $g$ satisfy conditions 1--3.
If $\tilde f'(g')$ is strictly convex (concave), then the maximum (minimum) $H_f$ for
fixed $H_g$ is obtained for probability distribution \eqref{maxer}, and the
minimum (maximum) $H_f$ is obtained for probability distribution \eqref{miner}.} \\

Here, and in the following derivations, the notation $\tilde f'(g')$ is equivalent to
$f'(\lambda(g'))$, and means $f'$ as a function of $g'$. Because the function $g$ is
strictly concave or convex, and the derivative $g'$ exists in the interval $(0,1)$,
$g'$ must be a one-to-one function of $\lambda$ in this interval. Hence it is possible to
invert this function to obtain $\lambda$ as a function of $g'$ (i.e.\ $\lambda(g')$).
In turn, we may express $f'$ as a function of $g'$ (i.e.\ $f'(\lambda(g'))$).

Note that the crucial relation the entropy measures must satisfy in order
for these bounds to hold is that $\tilde f'(g')$ is strictly concave or strictly
convex. The other restrictions on the functions $f$ and $g$ are
simply necessary to ensure that these are valid entropy measures.

It is also important to note that, for each value of $H_g$, the probability
distributions of the forms \eqref{maxer} and \eqref{miner} are unique, and hence
we obtain unique values for the upper and lower limits on $H_f$. Therefore the
bounds on $H_f$ obtained using this method are unambiguous.
To show this result for \eqref{maxer}, the value of $H_g$ is given by
\begin{equation}
H_g = g(\lambda_0) + (d-1) g[(1-\lambda_0)/(d-1)]
\end{equation}
so
\begin{equation}
\frac{\text{d}H_g}{\text{d}\lambda_0} = g'(\lambda_0) - g'[(1-\lambda_0)/(d-1)].
\end{equation}
Because $g'$ is one-to-one and $\lambda_0\ge (1-\lambda_0)/(d-1)$,
$\text{d}H_g/\text{d}\lambda_0$ always has the same sign, except
at $\lambda_0=(1-\lambda_0)/(d-1)$ where it is zero. This point is a boundary
to the range of $\lambda_0$; thus, $H_g$ must be a one-to-one function of
$\lambda_0$. Hence, for each value of $H_g$, there is a unique value of
$\lambda_0$ and therefore a unique probability distribution \eqref{maxer}.

The situation is similar for the probability distribution \eqref{miner}, except
there is an additional complication due to multiple values of $k$. In the same
way as for \eqref{maxer}, we can see that in each interval where $k$ is a
constant $(1/(k+1),1/k]$, $H_g$ is a one-to-one function of
$\lambda_0$. In particular, if $g$ is convex (concave),
then $H_g$ is monotonically increasing (decreasing). In addition, it is easy to
see that $H_g$ is continuous at the boundaries where $1/\lambda_0$ is an integer.
Thus $H_g$ is a one-to-one function of $\lambda_0$, and each value of $H_g$
corresponds to a unique probability distribution \eqref{miner}.

The method we will use for the proof is to first consider the restricted case
for three probabilities in lemmas 3--5, and then apply the result to prove theorem
1. When there are only three probabilities, the problem reduces to finding the
maximum and minimum of a function of a single real variable. This problem is
relatively straightforward, and may be solved by finding the boundaries of the
domain of the function, as well as the turning points.
Before we proceed to the case for three probabilities, there is a minor result
that we need to prove for the case of two probabilities. \nl
{\bf Lemma 2.} {\it Let $g:[0,1]\mapsto\mathbb{R}$ be a function satisfying conditions 1--3.
In addition, let $\lambda_0$ and $\lambda_1$ be two numbers in the interval
$[0,1]$, with the constraints
\begin{equation}
\label{lem2con}
\lambda_0+\lambda_1 = \Lambda^{(2)} \qquad g(\lambda_0)+g(\lambda_1) =H_g^{(2)}
\end{equation}
where $0\le\Lambda^{(2)}\le 1$. There are at most two solutions to \eqref{lem2con},
and these solutions differ by a permutation.} \hl

We use the notation convention that a superscript indicates a sum over fewer
than $d$ probabilities. The numbers $\lambda_0$ and $\lambda_1$ are only two
probabilities, so $H_g^{(2)}$ is not the same as the entropy $H_g$. The
superscript $(2)$ indicates that only two terms have been summed. \nl
{\bf Proof.} Solving \eqref{lem2con} is equivalent to solving
\begin{equation}
\label{h2}
g(\lambda_0)+g(\Lambda^{(2)}-\lambda_0)=H_g^{(2)}.
\end{equation}
If $\Lambda^{(2)}=0$, then there is only one solution, $\lambda_0=\lambda_1=0$.
If $\Lambda^{(2)}\ne 0$, then we may determine the number of solutions by
considering the turning points of the left-hand-side (LHS). For a turning point
we require $g'(\lambda_0)-g'(\Lambda^{(2)}-\lambda_0)=0$.
Since $g'$ is one-to-one, the only turning point is for
$\lambda_0=\Lambda^{(2)}/2$. Thus there can be at most two different values of
$\lambda_0$ that give the same value for the LHS of \eqref{h2}.
Denoting one solution for $\lambda_0$ as $\lambda_g$ (so
$\lambda_1=\Lambda^{(2)}-\lambda_g$), the other solution is for
$\lambda_0=\Lambda^{(2)}-\lambda_g$, in which case $\lambda_1=\lambda_g$. Therefore
the two solutions are simply related by a permutation. \hfill $\Box$ \hl

Next we require a result on the problem of finding the region of values that
the three probabilities may take given restrictions on these probabilities. \nl
{\bf Lemma 3.} {\it Let $g:[0,1]\mapsto\mathbb{R}$ be a function satisfying
conditions 1--3. In addition, let $\lambda_0$, $\lambda_1$ and $\lambda_2$ be
real numbers in the interval $[0,1]$ with the restrictions
\begin{equation} \label{rest1}
\lambda_0+\lambda_1+\lambda_2=\Lambda^{(3)}
\end{equation}
where $0\le\Lambda^{(3)}\le 1$ and
\begin{equation} \label{rest2}
g(\lambda_0)+g(\lambda_1)+g(\lambda_2)=H_g^{(3)}.
\end{equation}
The number $\lambda_0$ may take values within one or more subintervals of $[0,1]$;
at the boundaries of these subintervals, either one of the $\lambda_i$ is equal to
zero, or two are equal.} \nl
{\bf Proof.} Note first that two possible boundaries for $\lambda_0$ are at
0 and $\Lambda^{(3)}$ (if $\lambda_0=\Lambda^{(3)}$ then $\lambda_1=\lambda_2=0$).
To find other possible boundaries, consider solving for $\lambda_1$ and $\lambda_2$
for a given $\lambda_0$. The expression to solve may then be given as
\begin{equation}
g(\lambda_0)+g(\lambda_1)+g(\Lambda^{(3)}-\lambda_0-\lambda_1)=H_g^{(3)}.
\end{equation}
For a given $\lambda_0$, $\lambda_1$ takes values in the region
$[0,\Lambda^{(3)}-\lambda_0]$, and the LHS has a turning point at
$\lambda_1=(\Lambda^{(3)}-\lambda_0)/2$. For the points $\lambda_1=0$ and
$\lambda_1=\Lambda^{(3)}-\lambda_0$, there is at least one $\lambda_i$ which is
zero, whereas at $\lambda_1=(\Lambda^{(3)}-\lambda_0)/2$, $\lambda_1$ and
$\lambda_2$ are equal.

The three points $\{0,(\Lambda^{(3)}-\lambda_0)/2,\Lambda^{(3)}-\lambda_0\}$
are the three possible values of $\lambda_1$ where the LHS is at a maximum or a
minimum for a given $\lambda_0$. If there is no solution for $\lambda_1$, then
the maximum and minimum are either both above or both below $H_g^{(3)}$. On the
other hand, if there is a solution, then $H_g^{(3)}$ must be between the maximum
and minimum, or equal to one of these values\footnote{We use the convention that
the terminology `between' means not equal unless otherwise specified.}.
The maximum and minimum vary continuously with $\lambda_0$. Therefore, as we pass
from a region where there is a solution to a region where there is no solution,
either the maximum or the minimum must pass through $H_g^{(3)}$.
Hence, at a boundary of the region of values that $\lambda_0$ may take,
either at least one of the $\lambda_i$ is zero, or at least two are equal.
\hfill $\Box$ \hl

Now we apply this result to the bounds problem for the case of three probabilities: \nl
{\bf Lemma 4.} {\it Let $f$ and $g$ be functions $[0,1]\mapsto\mathbb{R}$ such that
conditions 1--3 are satisfied. Let $\lambda_0$, $\lambda_1$, and $\lambda_2$ be
real numbers in the interval $[0,1]$ with the restrictions \eqref{rest1} and
\eqref{rest2}, and let $H_f^{(3)}$ be defined by
\begin{equation}
\label{lem4eq}
H_f^{(3)}\equiv f(\lambda_0)+f(\lambda_1)+f(\lambda_2).
\end{equation}
If $\tilde f'(g')$ is strictly convex or concave, then the extremal values of $H_f^{(3)}$
are obtained when at least one of the $\lambda_i$ is zero or at least two are
equal.} \nl
{\bf Proof.} Note first that, from lemma 2, for a given value of $\lambda_0$
there are at most two solutions of \eqref{rest1} and \eqref{rest2} for
$\lambda_1$ and $\lambda_2$, and one solution is a permutation of the other.
Thus, for each value of $\lambda_0$, there is only one possible value of
$H_f^{(3)}$.
Therefore $H_f^{(3)}$ is a single-valued function of $\lambda_0$, and we may
find the maximum and minimum by finding the boundaries of the region of values
that $\lambda_0$ may take, as well as the turning points. From lemma 3, at the
boundaries of the region of values that $\lambda_0$ may take, either one of the
$\lambda_i$ is zero, or at least two are equal.

To complete the proof, it remains to be shown that there are no turning points
for values of $\lambda_0$ such that the $\lambda_i$ are nonzero and unequal.
For a turning point, we require that $\text{d}H_f^{(3)}/\text{d}\lambda_0$
changes sign. Taking the derivative of \eqref{lem4eq} with respect to
$\lambda_0$ gives
\begin{equation} \label{Sder0}
\frac{\text{d}H_f^{(3)}}{\text{d}\lambda_0}=f'(\lambda_0)+\frac{\text{d}
\lambda_1}{\text{d}\lambda_0}f'(\lambda_1)+\frac{\text{d}\lambda_2}{\text{d}
\lambda_0}f'(\lambda_2).
\end{equation}
To remove the derivatives $\text{d}\lambda_1/\text{d}\lambda_0$ and
$\text{d}\lambda_2/\text{d}\lambda_0$, we may take the derivatives of
\eqref{rest1} and \eqref{rest2} with respect to $\lambda_0$, and substitute
into \eqref{Sder0}. We then obtain
\begin{align}
\label{Sder}
\frac{\text{d}H_f^{(3)}}{\text{d}\lambda_0}&=\left[ \frac{f'(\lambda_1)-
f'(\lambda_0)}{g'(\lambda_1)-g'(\lambda_0)}-\frac{f'(\lambda_2)-f'(\lambda_0)}
{g'(\lambda_2)-g'(\lambda_0)}\right] 
\left\{\frac{[g'(\lambda_2)-g'(\lambda_0)]
[g'(\lambda_1)-g'(\lambda_0)]}{g'(\lambda_1)-g'(\lambda_2)}\right\}.
\end{align}
Because $g'$ is one-to-one, if the $\lambda_i$ are unequal, then
the terms in the denominators are nonzero, and the derivative
${\text{d}H_f^{(3)}}/{\text{d}\lambda_0}$ is continuous. In that case, for there
to be a turning point, we require that the derivative is zero, which implies
\begin{equation}
\frac{f'(\lambda_1)-f'(\lambda_0)}{g'(\lambda_1)-g'(\lambda_0)}=
\frac{f'(\lambda_2)-f'(\lambda_0)}{g'(\lambda_2)-g'(\lambda_0)}.
\end{equation}
This expression implies that the three points $(g'(\lambda_0),f'(\lambda_0))$,
$(g'(\lambda_1),f'(\lambda_1))$ and $(g'(\lambda_2),f'(\lambda_2))$ lie along
a straight line. This is not possible with unequal probabilities if $\tilde f'(g')$ is
strictly convex or strictly concave. Thus we see that
there are only two possibilities for a maximum or minimum of $H_f^{(3)}$: one
of the $\lambda_i$ is zero, or two are equal. \hfill $\Box$ \hl

The last lemma we show is a refinement of lemma 4 to account for when the various
solutions occur and whether they give a maximum or a minimum. \nl
{\bf Lemma 5.} {\it Let $f$ and $g$ be functions $[0,1]\mapsto\mathbb{R}$ such that
conditions 1--3 are satisfied. Let $\lambda_0$, $\lambda_1$ and $\lambda_2$ be
real numbers in the interval $[0,1]$ with the restrictions \eqref{rest1} and
\eqref{rest2}, and let $H_f^{(3)}$ be defined as in \eqref{lem4eq}.
If $\tilde f'(g')$ is strictly convex (concave),
then the maximum (minimum) $H_f^{(3)}$ is obtained only if two $\lambda_i$ are
equal and one is larger or equal, and the minimum (maximum) $H_f$ is obtained
only if one of the $\lambda_i$ is zero, or two are equal and one is smaller
or equal.} \nl
{\bf Proof.} In the case that one of the $\lambda_i$ is zero, we may take
$\lambda_0$ to be zero without loss of generality. Using lemma 1, the extremal
values of $H_g^{(3)}$ are $g(\Lambda^{(3)})$ and $2g(\Lambda^{(3)}/2)$.
In general, the extremal values of $H_g^{(3)}$ are obtained for all $\lambda_i$
equal, giving $H_g^{(3)}=3g(\Lambda^{(3)}/3)$, and all probabilities zero except
for one, giving $H_g^{(3)}=g(\Lambda^{(3)})$. Therefore $H_g^{(3)}$ may take
values from $g(\Lambda^{(3)})$ to $3g(\Lambda^{(3)}/3)$. It is easily seen that
$2g(\Lambda^{(3)}/2)$ lies in this interval. If $H_g^{(3)}$ lies between
$g(\Lambda^{(3)})$ and $2g(\Lambda^{(3)}/2)$, or is equal to one of these values,
then there is a solution with $\lambda_0=0$. If $H_g^{(3)}$ is between
$2g(\Lambda^{(3)}/2)$ and $3g(\Lambda^{(3)}/3)$, or equal to $3g(\Lambda^{(3)}/3)$,
there is no solution with $\lambda_0=0$.

For the other case, where at least two of the $\lambda_i$ are equal, we
may take $\lambda_1=\lambda_2$ without loss of generality. Then the restrictions
\eqref{rest1} and \eqref{rest2} give
\begin{equation}
\label{eqeq}
g(\lambda_0)+2g[(\Lambda^{(3)}-\lambda_0)/2]=H_g^{(3)}.
\end{equation}
To determine the number of solutions to this, we may consider the LHS as a
function of $\lambda_0$. The number $\lambda_0$ is in the range
$[0,\Lambda^{(3)}]$, and there is one turning point at $\lambda_0=\Lambda^{(3)}/3$.
Therefore there may be at most two solutions to equation \eqref{eqeq}.

\begin{figure}
\begin{center}
\includegraphics[width=0.5\textwidth]{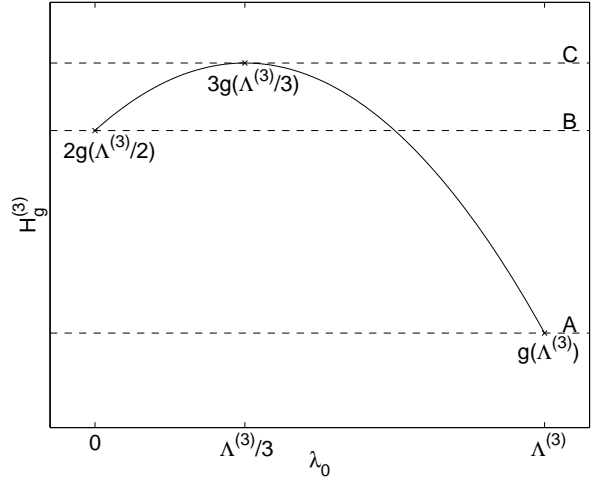}
\end{center}
\caption{An example of the variation of the LHS of equation \eqref{eqeq} as a
function of $\lambda_0$ for the case where $g$ is strictly concave.}
\label{illust}
\end{figure}

To be more specific, the LHS takes the values $2g(\Lambda^{(3)}/2)$,
$3g(\Lambda^{(3)}/3)$ and $g(\Lambda^{(3)})$ for $\lambda_0$ equal to $0$,
$\Lambda^{(3)}/3$ and $\Lambda^{(3)}$, respectively. This situation is illustrated in
figure \ref{illust} for the case where $g$ is strictly concave. There are
five qualitatively different situations for different values of
$H_g^{(3)}$ from $g(\Lambda^{(3)})$ to $3g(\Lambda^{(3)}/3)$:
\newcounter{Lcount}
\begin{list}{\arabic{Lcount}.}
{\usecounter{Lcount}}
\item For $H_g^{(3)}=g(\Lambda^{(3)})$ (line A), there is only one possible
solution, which corresponds to all the $\lambda_i$ being zero except one.
\item If $H_g^{(3)}$ lies between $g(\Lambda^{(3)})$ and $2g(\Lambda^{(3)}/2)$ (between
lines A and B), there can be only one solution with $\lambda_1=\lambda_2$. This
solution is for $\lambda_0$ in the interval $(\Lambda^{(3)}/3,\Lambda^{(3)})$.
There will also be a solution with $\lambda_0=0$, giving a total of two solutions.
\item If $H_g^{(3)}$ is equal to $2g(\Lambda^{(3)}/2)$ (line B), then there are two
solutions with $\lambda_1=\lambda_2$, one for $\lambda_0=0$ and the other for
$\lambda_0$ in the interval $(\Lambda^{(3)}/3,\Lambda^{(3)})$. We can also obtain a
solution by setting $\lambda_0=0$; however, this solution is identical to the
solution for $\lambda_1=\lambda_2$ where $\lambda_0=0$. Therefore there are only
two distinct solutions.
\item For $H_g^{(3)}$ in the range between $2g(\Lambda^{(3)}/2)$ and
$3g(\Lambda^{(3)}/3)$ (between lines B and C), there are two solutions with
$\lambda_1=\lambda_2$, one for $\lambda_0$ in the range $(0,\Lambda^{(3)}/3)$ and
the other for $\lambda_0$ in the range $(\Lambda^{(3)}/3,\Lambda^{(3)})$. There is no
solution with $\lambda_0=0$, again resulting in a total of two solutions.
\item The last possibility is $H_g^{(3)}=3g(\Lambda^{(3)}/3)$ (line C), in which case
all the $\lambda_i$ are equal.
\end{list}

Thus we find that, for $H_g^{(3)}$ between $3g(\Lambda^{(3)}/3)$ and
$g(\Lambda^{(3)})$, there are always two solutions where either one of the
$\lambda_i$ is zero or two are equal, thereby providing a maximum and minimum for
$H_f^{(3)}$. If $H_g^{(3)}$ is equal to $3g(\Lambda^{(3)}/3)$ or $g(\Lambda^{(3)})$,
there is only one possible solution, so the maximum and minimum coincide.

In order to determine which solution gives the maximum of $H_f^{(3)}$ and which
gives the minimum, let us consider the solution where $\lambda_0>\Lambda^{(3)}/3$
and $\lambda_1=\lambda_2$. For $H_g^{(3)}$ between $g(\Lambda^{(3)})$ and
$3g(\Lambda^{(3)}/3)$ there are always two distinct solutions, one of which is of
this form. Therefore we may determine which solution gives the minimum and which
gives the maximum by only considering this case. This value of $\lambda_0$ is an
upper boundary, because the other solutions for $\lambda_0$ are smaller.
Let us consider a value of $\lambda_0$ slightly below this solution, so
$\lambda_1\ne\lambda_2$. We may take $\lambda_1$ to be the larger value, so
$\lambda_0>\lambda_1>\lambda_2$.

If $g$ is convex, then $g'(\lambda_0)>g'(\lambda_1)>g'(\lambda_2)$, and if $g$ is
concave, then $g'(\lambda_0)<g'(\lambda_1)<g'(\lambda_2)$. Therefore the
multiplying factor in braces in equation \eqref{Sder} is positive if $g$ is convex,
and negative if $g$ is concave. It is also easy to see that, if $\tilde f'(g')$ and $g$
are both convex or both concave, then the first term in the square brackets in
equation \eqref{Sder} is greater than the second term. If one of $\tilde f'(g')$ and $g$
is convex and the other is concave, then the first term in the square brackets is
smaller than the second term.

Thus we find that ${\text{d}H_f^{(3)}}/{\text{d}\lambda_0}$ is positive if
$\tilde f'(g')$ is convex, and negative if $f'(g')$ is concave. Therefore, for
$\tilde f'(g')$ convex, as we increase $\lambda_0$ up to its maximum value, $H_f^{(3)}$
is increasing, and the solution where $\lambda_0>\lambda_1=\lambda_2$ must be a
maximum. Similarly, for $\tilde f'(g')$ concave, the solution where
$\lambda_0>\lambda_1=\lambda_2$ is a minimum. \hfill $\Box$ \hl

Now that we have solved the case for three probabilities, we may extend the
solution to the general case with $d$ probabilities. Thus the proof of theorem
1 is as given below. \nl
{\bf Proof of theorem 1.} We wish to find the maximum and minimum
of $H_f=\sum_i f(\lambda_i)$ with fixed $H_g=\sum_i g(\lambda_i)$ and
$\sum_i\lambda_i=1$. To solve this case, let $\{\lambda_i\}$ be a set of
probabilities that maximises $H_f$. We then select any three probabilities
$\lambda_{i_0}$, $\lambda_{i_1}$ and $\lambda_{i_2}$, and define $I$ to be the
set of indices $\{i_0,i_1,i_2\}$, and $I^\perp$ to be the set of indices
excluding $I$. If the set of probabilities $\{\lambda_i\}$ maximises the sum
$\sum_i f(\lambda_i)$, then the set $\{\lambda_{i_0},\lambda_{i_1},
\lambda_{i_2}\}$ must maximise the sum
\begin{equation}
H_f^I = \sum_{i\in I} f(\lambda_i)
\end{equation}
with the restrictions
\begin{equation}
\sum_{i\in I} \lambda_i = \Lambda^{(3)} \qquad \sum_{i\in I} g(\lambda_i) = H_g^I
\end{equation}
where $\Lambda^{(3)} = 1-\sum_{i\in I^\perp} \lambda_i$ and $H_g^I = H_g -
\sum_{i\in I^\perp}g(\lambda_i)$. From lemma 5, if $\tilde f'(g')$ is
strictly convex (concave), $H_f^I$ is maximised (minimised) {\it only} with
probabilities of the form $\lambda_0\ge\lambda_1=\lambda_2$. That is, $H_f^I$
being maximised (minimised) implies that the probabilities are of this form. The
only way that this criterion can be satisfied for all subsets of three
probabilities is if all the probabilities are equal, except for one which may be
larger. Therefore the probability distribution must be of the form \eqref{maxer}.

Similarly, for $\tilde f'(g')$ strictly convex (concave), $H_f^I$ will be
minimised (maximised) only for probabilities of the form
$\lambda_0\le\lambda_1=\lambda_2$, or for one of the probabilities equal to zero.
The only way that this criterion can be satisfied for all subsets of three
probabilities is if a number of the probabilities are equal, one is smaller and
the rest are zero. Therefore the probability distribution must be of the form
\eqref{miner}. Thus we have proven each of the alternative cases for theorem 1.
\hfill $\Box$ \hl

Although we have presented the above analysis in terms of probabilities,
identical results hold for entropies of mixed states and entanglements of pure
states. The eigenvalues of density matrices or Schmidt coefficients of pure
entangled states may be analysed in the same way as probabilities, so
we have the following two corollaries. \nl
{\bf Corollary 1.} {\it Let $S_g(\rho)$ and $S_f(\rho)$ be two entropy measures where
the functions $f$ and $g$ satisfy conditions 1--3. If $\tilde f'(g')$ is strictly
convex (concave), then the maximum (minimum) $S_f$ for fixed $S_g$ is obtained
for a state with eigenvalues of the form \eqref{maxer}, and the minimum (maximum)
$S_f$ is obtained for a state with eigenvalues of the form \eqref{miner}.} \nl
{\bf Corollary 2.} {\it Let $E_g(\ket{\psi})$ and $E_f(\ket{\psi})$ be two entanglement
measures where the functions $f$ and $g$ satisfy conditions 1--3. If $\tilde f'(g')$
is strictly convex (concave), then the maximum (minimum) $E_f$ for fixed $E_g$ is
obtained for a state with Schmidt coefficients of the form \eqref{maxer}, and the
minimum (maximum) $E_f$ is obtained for a state with Schmidt coefficients of the
form \eqref{miner}.}

\section{Applications}
\label{sec:apply}
Next we consider applications of these results. As our first application,
consider the problem of maximising or minimising the von Neumann entropy for
given linear entropy. In this situation we take
\begin{equation}
f(\lambda) = -\lambda\log\lambda \qquad g(\lambda)=\lambda^2.
\end{equation}
Both $f$ and $g$ satisfy conditions 1--3. We find that
\begin{equation}
f'(\lambda) = -\log\lambda-\log e \qquad g'(\lambda)=2\lambda.
\end{equation}

To determine if $\tilde f'(g')$ is strictly convex or strictly concave, we may calculate
$\text{d}\tilde f'/\text{d}g'$. When this derivative is monotonically increasing
(decreasing), $\tilde f'(g')$ is strictly convex (concave). We may determine
$\text{d}\tilde f'/\text{d}g'$ from $f''/g''$, which gives
\begin{equation}
\frac{f''}{g''} = -\frac {\log e}{2\lambda}.
\end{equation}
Thus $\text{d}\tilde f'/\text{d}g'$ is monotonically increasing, and $\tilde f'(g')$ is
strictly convex. This implies, from corollary 1, that the von Neumann entropy is
maximised for a density matrix with eigenvalues of the form \eqref{maxer}, and
minimised when the eigenvalues are of the form \eqref{miner}. Therefore, for the
case of the von Neumann entropy and the linear entropy, we obtain the result
given in \cite{Harre,Wei}.

As another application, we may consider the comparison of two $\alpha$ entropies
for different values of $\alpha$:
\begin{equation}
f(\lambda)=\lambda^{\alpha_1} \qquad
g(\lambda)=\lambda^{\alpha_2}
\end{equation}
where $\alpha_1\ne\alpha_2$. For this example we obtain
\begin{equation}
\frac{f''}{g''} = \frac {\alpha_1(\alpha_1-1)}
{\alpha_2(\alpha_2-1)}\lambda^{\alpha_1-\alpha_2}.
\end{equation}
This expression will be either monotonically increasing or monotonically
decreasing depending on the values of $\alpha$. Therefore the maximum and minimum
$\alpha_1$ entropies are again obtained for the same form of states. Note that,
because taking $\alpha\to 1$ or 2 also gives the von Neumann entropy and the linear
entropy, these bounds hold for any two-way comparison between these entropy measures.

As an example of a case where the maximum and minimum entropies are not given by
states with eigenvalues of the form \eqref{maxer} and \eqref{miner}, consider
\begin{equation}
\label{peculiar}
f(\lambda)=-\lambda\log\lambda \qquad g(\lambda)=\lambda^2+
1.99[1-\cos(\omega\lambda)]/\omega^2
\end{equation}
where we take $\omega=10$. The entropy $S_f$ is simply the von Neumann
entropy, whereas $S_g$ is slightly modified from the purity.
We find that $g''(\lambda)=2+1.99\cos(\omega\lambda)$,
which does not change sign, so $S_g$ is a valid entropy measure. However,
\begin{equation}
\frac{f''}{g''} = -\frac {\log e}
{\lambda[2+1.99\cos(\omega\lambda)]}
\end{equation}
is not one-to-one for $\omega=10$. Therefore the entropies $S_f$ and $S_g$ do
not satisfy the conditions of theorem 1, even though they are valid entropy
measures.

\begin{figure}
\begin{center}
\includegraphics[width=0.475\textwidth]{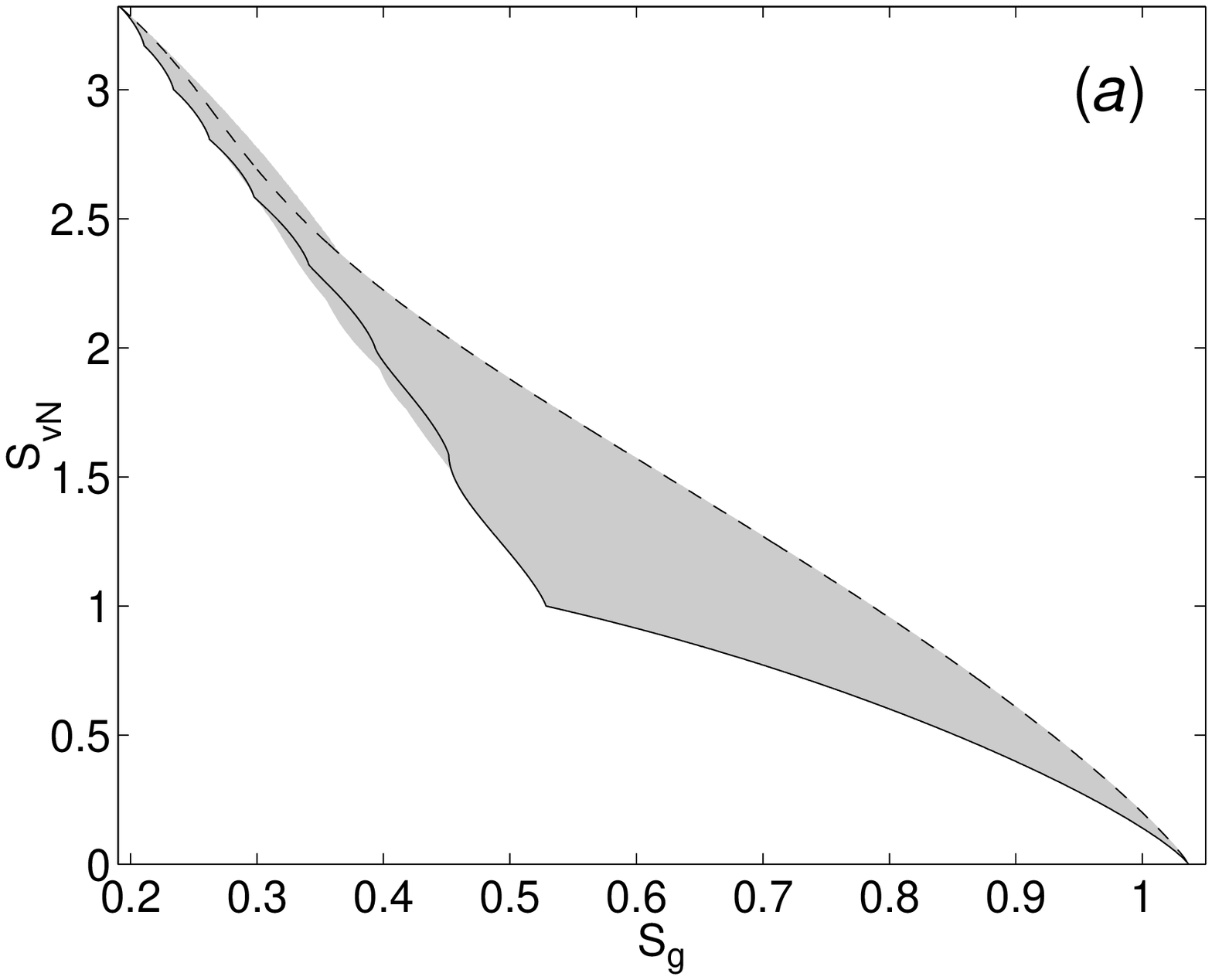}
\includegraphics[width=0.48\textwidth]{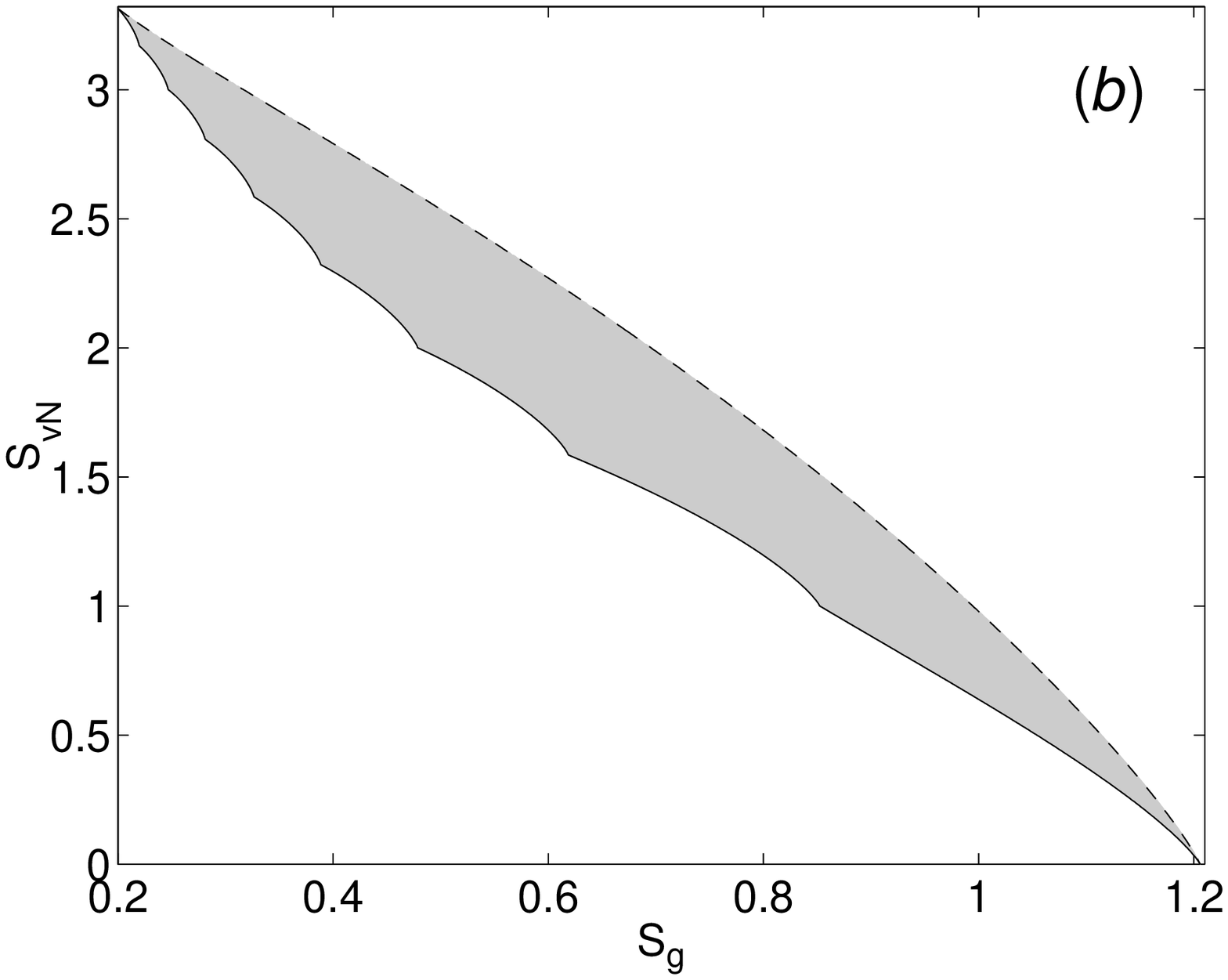}
\end{center}
\caption{The von Neumann entropy versus the
$S_g$ entropy with $g(\lambda)$ given by equation \eqref{peculiar} with
$\omega=10$ ($a$) and $\omega=4$ ($b$). The limit that would be given by states of
the form \eqref{miner} is shown by the solid line and the limit that would be
given by states of the form \eqref{maxer} is shown by the dashed line. The
shaded region is composed of a large number of points for randomly generated
states. The dimension is $d=10$.}
\label{counter}
\end{figure}

The limits that would be given by states with eigenvalues of the forms
\eqref{maxer} and \eqref{miner} are shown in figure \ref{counter}($a$). In
addition, results for a large number of randomly generated states are shown. A
number of points lie outside the boundaries given by the states of the form
\eqref{maxer} and \eqref{miner}, demonstrating that these do not provide the
limits to the von Neumann entropy for given $S_g$ entropy.

Nevertheless, even for this example, most of the points lie within the region
between the two curves, and the points that lie beyond the boundary are only a
small distance from the boundary. This example has been chosen because the
points are at a noticeable distance from the boundaries. For other values of
$\omega$ the difference is not so noticeable. In fact, for some values of
$\omega$ there were no points found to lie beyond the boundary, despite the fact
that $\tilde f'(g')$ is not strictly convex or concave. An example for $\omega=4$ is
shown in figure \ref{counter}($b$). These results strongly indicate that the
condition that $\tilde f'(g')$ is strictly convex or concave is not a necessary
condition, although it is sufficient.

\section{Conclusions}
\label{sec:conclude}
We have shown how to determine the maximum and minimum possible values of one
type of entropy for fixed values of another type of entropy. The forms of states
that achieve these maximum and minimum values are the same as those given by
\cite{Harre,Wei} for the case of comparing the von Neumann entropy
to the linear entropy. These results may be applied to entropies of probability
distributions, entropies of mixed states and measures of entanglement for pure
states.

We have identified the relation between the entropy measures that is
necessary for these bounds to hold. This relation holds between the von Neumann
entropy, the linear entropy and the $\alpha$ entropy. The bounds we have derived
therefore apply to any two-way comparison between these entropy measures. These
results allow one to estimate the value of one type of entropy given the value
of another.

For the examples we have examined, we have found that these bounds are a good
approximation of the true bounds even for comparisons between entropy measures
that do not satisfy the conditions of the proof. This indicates that, even in
such cases, the bounds we have found may be used to estimate the value of one
entropy from the other (without giving the exact bounds).

\section*{Acknowledgments}
The authors are grateful to William Munro, who helped with the manuscript,
and Karol \.{Z}yczkowski, who shared a preliminary version of his manuscript
with us. The authors also acknowledge valuable discussions with Xiaoguang
Wang, Stephen Bartlett and Robert Spekkens. This research has been supported by
an Australian Research Council Large Grant and by a Department of Education
Science and Training Innovation Access Program Grant to support the European
Fifth Framework Project QUPRODIS.

\end{document}